# Relativistic Aberration Effect in Astrophysics and Cosmology

O. G. Semyonov[1]

*State University of New York at Stony Brook, 231 Old Chemistry, Stony Brook, NY 11794, USA*

ABSTRACT

Relativistic aberration influences apparent brightness/luminosities of objects moving with relativistic relative velocities. The superluminosity or dimming of incoming/receding jets ejected from Active Galactic Nuclei is believed to be the manifestation of the effect. Redshifted cosmological objects, such as high-z galaxies and supernovae in high-z galaxies, are also subjected to luminosity dimming due to relativistic aberration, although the correcting terms encompassing spectral redshift and time dilation are used to scale the observed magnitudes in the restframe template. The Universe's expansion results in elongation of distances to the objects from the moment of emission to the present time related to the z = 0 local space; this elongation is equal to the apparent increase of the *distance from the apparent size* to the restframe distance in the theory of special relativity due to the effect of relativistic aberration. Contrary to the standard Big Bang model with stationary space, the effect of relativistic aberration zeroes in the expending space model; if local spaces and local objects expand together with the Universe, the luminosity distances to high-redshifted objects appear to be 1+z times smaller then in the Big Bang model.

1. INTRODUCTION

Relativistic aberration of light is the cause of shift of apparent angular positions of objects on the celestial sphere due to the motion of Earth around our Sun (Pauli 1958) and the motion of Sun around the center of Galaxy (Klioner & Soffel 1998). Less known is the searchlight effect that displays the impact of relativistic aberration on the apparent brightness/luminosity of relativistic objects, such as jets ejected from Active Galactic Nuclei (AGN) (e.g., Biretta et al. 1999). As for the redshifted cosmological objects, their apparent brightness and luminosity are believed to be uninfluenced by the relativistic aberration; this corollary, being untrue in essence, is nevertheless justified because their coordinates, redshifted due to the Universe's expansion, are referred at the moment of detection to the restframe z = 0. In the online-published paper (Semyonov 2004), it was attempted to explain the dimming of high-z supernovae by the effect of relativistic aberration, however this conclusion, as shown below, is mistaken.

---

[1] osemyonov@ece.sunysb.edu

The evaluation of luminosity distances to the cosmic objects is of great importance for verifying the cosmological models. While the measurement of redshift z is based on observations of shifts of spectral lines whose original spectral positions in any local space are determined by the atomic constants that seem to be the same at any part of the Universe, the evaluation of distances to remote extra-galactic objects is not so straightforward. The main problem lies firstly in finding the standard 'candles' of constant intrinsic luminosity that does not depend on their actual distance from us and secondly in taking account of the factors that can influence their apparent luminosity. It is assumed that radiation flux $F$ of a luminous pointlike object is a function of its luminosity L (energy per steradian per unit time) and distance between the object and our detector D:

$$F = \frac{L}{D^2}, \qquad (1)$$

Therefore if one can select the standard 'candles' having the same intrinsic luminosities L, then (1) can be solved for D provided the observed flux $F$ (magnitude m) is known from the measurements. The commonly used correcting terms for the apparent magnitudes of deep-space objects include: relativistic redshift, time dilation if applicable (Schmidt 1998), light absorption in our Galaxy and in host galaxies, evolution of galaxies in the expending Universe where the redshift z is also associated with the age of the observed galaxy (Humason et al. 1956, Sandage 1995), and gravitational lensing. The redshift-related corrections of magnitudes are commonly include the K-correction term, which encompasses bandwidth and bandshift of galaxies (Rowan-Robinson 1981, Sandage 1995), clusters of galaxies (Collins and Mann 2004, Reichart et al. 1999), and supernovae (Schmidt 1998; Riess 1998; Perlmutter 1999), and the ∆-correction term, encompassing time dilation to transform the light curves of supernovae to the restframe z = 0 (Schmidt 1998; Riess 1998; Perlmutter 1999).

In this paper, the searchlight effect caused by relativistic aberration of light emitted from relativistic sources is analyzed and its applicability to the redshifted cosmological objects is discussed in the context of their luminosity distances and apparent sizes.

## 2. RELATIVISTIC ABERRATION IN SPECIAL RELATIVITY

The effect of light aberration (Pauli 1958; Møller 1972) of moving relativistic sources is related to apparent deflection of light rays in the observer's space with respect to the comoving frame. Consider two coordinate systems S and S´ originated from a poinlike isotropic source at an arbitrary moment of time as shown in Fig. 1. The system S is at rest with respect to a detector and the system S´ comoves together with the source with the velocity β = v/c along the z-axis in positive direction. A light beam that makes an angle θ´ with the velocity vector (z-axis) in the frame S´ co-moving with the source will have an apparent direction in the reference frame S, which is characterized by another angle θ, and θ´ and θ are linked by (Møller 1972):

$$\cos\theta' = \frac{\cos\theta - \beta}{1 - \beta\cos\theta}, \qquad (2)$$

with the azimuth angles φ = φ´ in the corresponding spherical coordinates. This formula displays, in particular, the searchlight effect, i.e. beaming of radiation in the direction of movement when it is observed from the unmoving detector's frame (for example, θ = arccos β < π/2 when θ´ = π/2). From the point of view of an observer in the frame S, the solid angles made by the beams emitted from the source are compressed with respect to the co-moving (restframe) coordinate system in the hemisphere turned towards the direction of movement and stretched out in the opposite hemisphere (Fig. 1b).

The searchlight effect is well-known in physics; it is observed, in particular, for synchrotron radiation of relativistic electrons moving in magnetic fields, for impact bremsstrahlung of relativistic electrons incident on a target at rest, and for products of nuclear reactions of relativistic particles with particles at rest. Resent observations of jets emitted by the Active Galactic Nuclei (AGN) also displayed the strong evidence of beaming of radiation (Biretta et al. 1999) and superluminosity/dimming of the incoming/receding jets. Bearing in mind the question of applicability of the effect to the redshifted cosmological objects, the further discussion will be focused on the receding sources.

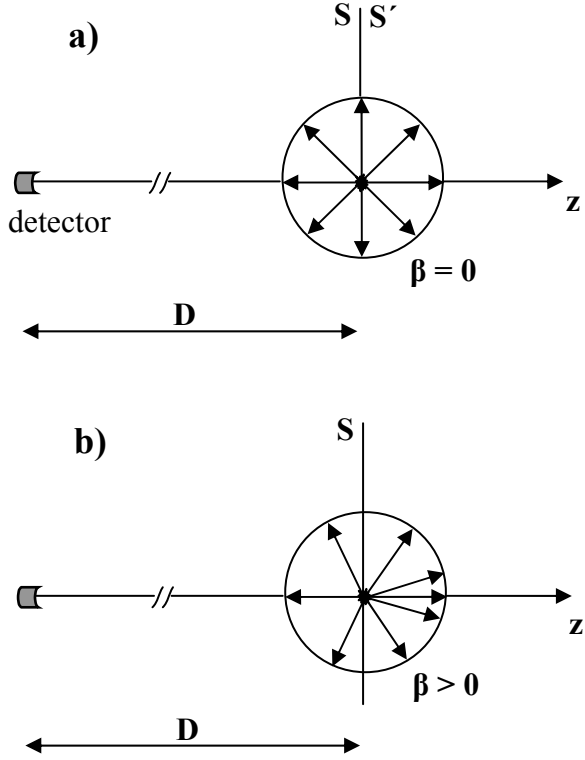

**Fig. 1: a)** Angular distribution of light rays emitted by an immobile isotropic source in the co-moving coordinate restframe S´; **b)** angular distribution of the same rays as seen from the coordinate frame S: the rays in the right hemisphere are inclined toward the z-axis (direction of movement) compressing the solid angles made by corresponding light rays and boosting the source luminosity (number of photons per steradian) while the rays in the left hemisphere decline out of z-axis stretching the solid angles and diminishing the source luminosity in the direction to detector. D is the actual distance from the detector to the source in the restframe. The sketch is drafted to demonstrate the phenomenon of luminosity change in the detector's frame S due to relativistic aberration and does not reflect other relativistic effects.

Following the deduction of McKinley (1980) and Rindler (1960, 50), it can be shown from the equation (2) that a backward-directed small solid angle made by light rays originated from a pointlike source moving away from a detector along the line of sight in a flat space is transformed with respect to the co-moving coordinate system S´ as $d\Omega = \delta^2 \, d\Omega' = d\Omega'/[\gamma^2(1 - \beta)^2] = d\Omega' \, (1+z)^2$, where $\gamma = (1 - \beta^2)^{-1}$ and z is the redshift. The effect results in reduction of luminosity (energy emitted per unit solid angle per unit time) by a factor $(1+z)^{-2}$ in the coordinate frame S because the photons emitted into a solid angle $d\Omega'$ in the restframe S´ propagate into a larger solid angle in the frame S. The outcome of the beaming effect is that the flux $F$ incident on a detector from a pointlike relativistic source with its intrinsic luminosity $L$ is diminished by a factor $(1+z)^{-2}$ due to reduced number of detected photons per unit area at any distance from the source in the coordinate frame S.

Consider an object that is seen as a luminous disk of visible area A with its intrinsic bolometric brightness $b$ in the restframe S′ (Fig. 2) observed from a distance D′. Its brightness is a characteristic of a point on the disk's surface emitting the isotropic radiation in all directions into the left hemisphere, therefore the relativistic aberration effect results in deflection of rays emitted from every point the same way as for a pointlike object; a couple of rays are shown in Fig. 2 for an arbitrary point by two solid arrows. The effect of relativistic aberration is nothing else but the apparent angular deflection of light rays emitted from every point of the receding disk and, in accordance with the expression (1), the apparent direction of the same rays in the frame S will be seen as shown in Fig. 2a by two dashed arrows.

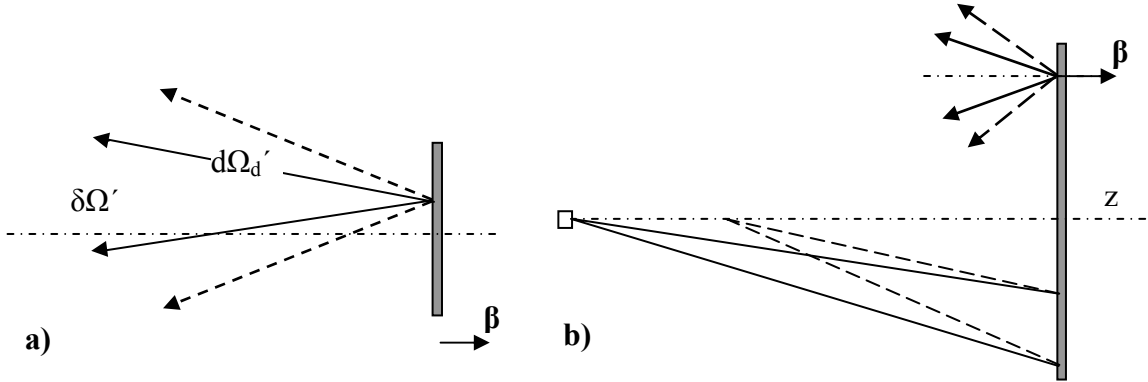

**Fig. 2. a)** Apparent deflection of rays emitted by a point on a moving-away disk (dashed arrows) with respect to their direction in restframe (solid arrows); **b)** Apparent 'focusing' of rays due to the effect of relativistic aberration (see explanations in the text).

The effect of deflection of rays results in the apparent widening of solid angles originated from any luminous point on the disk's surface with respect to the restframe, i.e. the photons radiated into a solid angle $\delta\Omega′$ in the restframe S′ are seen as emitted into a solid angle $\delta\Omega′(1+z)^2$ in the frame S. It means that the brightness of the disk (radiation power per unit area per steradian) observed from the frame S along the direction of the disk's regress is reduced by a factor $(1+z)^{-2}$ due to reduced number of photons emitted into unit solid angle.

The aberrational deflection of rays emitted from every point on the disk's surface produces the effect of 'focusing' at some point positioned on the axis closer to the disk, as shown in Fig. 2b. Let's choose a z-axis drawn through our detector and the center of the luminous disk perpendicular to its area. Every point on its surface radiates isotropic radiation in the restframe where the disk is stationary and the rays are distributed isotropically over the angles. Select two of them originated from an arbitrary point on the disk surface as shown in Fig. 2b (solid arrows). The effect of relativistic aberration of light emitted by a receding disk results in deflection of all the rays further from the axis collinear with the velocity vector of this point (dashed arrows). One can notice that the rays directed out of our chosen z-axis are deflected further from this z-axis and the rays inclined to this axis incline to it even more. Our detector positioned at a distance D´ in the frame S´ sees only the rays inclined to the axis from any luminous point outside the disk's center; the inclination due to relativistic aberration results in apparent 'focusing' of deflected rays to a point on the axis located closer to the disk, as it is shown for a couple of separate points on the disk's surface below the axis. The effect enlarges the apparent angular size of the disk (visibility angle of the source from the detector's location) with respect to the solid angle $d\Omega_d´ = A/D´^2$ subtended by the disk in the restframe S´ (Fig. 3), therefore the apparent distance between the disk and the detector looks shorter in the frame S. Following Rindler (1960), the apparent distance D to the extended object deducted from its apparent angular size $d\Omega_d = A/D^2$ in the frame S is:

$$D = \left(\frac{A}{d\Omega_d}\right)^{1/2} = \left(\frac{A}{d\Omega_d'}\right)^{1/2} \left(\frac{d\Omega_d'}{d\Omega_d}\right)^{1/2} = \frac{D'}{1+z}.$$

This is the *distance from the apparent size*: the disk looks (1+z) times closer to the detector in comparison with the distance to it in the restframe because the solid angle, it subtends, is increased by a factor $(1+z)^2$. It can be shown that the distance from the apparent size is the distance to the disk at the moment of emission of now-detected photons. Consider a spherical wavefront initiated in space by a short event of emission from a point on the disk when it was at a distance D from a detector. In the frame S the wavefront is also spherical due to constancy of speed of light *c*, however it always stays centered on a point in space located at the same distance D, while the disk moves further away from the point of emission. This wavefront reaches the detector in a time interval D/*c* after emission which is (1+z) times shorter than it would be in the frame S´. At the moment of detection the disk's actual position D(1+z) in the frame S becomes equal to the distance D´ in the frame S´.

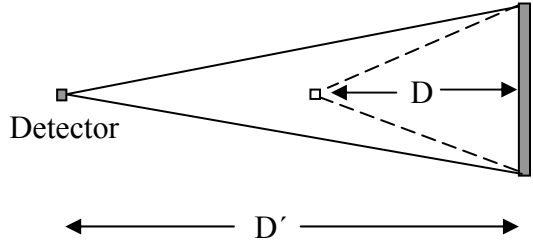

Fig. 3. Visible angular size of a luminous disk at rest (solid lines) and moving away from the detector (dashed lines).

For a seemingly closer object, the radiation flux at the detector's location captured from a finite solid angle $d\Omega_d$ on the sky is $(1+z)^2$ higher in comparison with the restframe; this increase of flux exactly cancels the brightness dimming due to the searchlight effect, if a detector (e.g., a radio telescope) with its solid angle of detection equal to the apparent angular size of the object is used for measuring the radiation flux. It does not mean that the searchlight effect vanishes; it works as usual and reduces the apparent brightness of the disk's surface. Brightness is an intrinsic characteristic of a light source obeying the relativistic analog of brightness conservation $b = b'/(1+z)^4$, where $b$ is the spectral brightness in the frame S and $b'$ is the spectral brightness in the restframe.

For an objective lens (or a mirror) of an imager all celestial sources are at infinity and the images are generated in a focal plane. In an optical system with focal length f, the area of image $A_i$ of an object subtending a small solid angle $d\Omega_d$ is $A_i = f^2 \, d\Omega_d$, therefore the image of object of larger angular size is correspondingly larger. The area of image of a receding object is $(1+z)^2$ times larger then it would be in the restframe, therefore the illuminance of the image $I = F A_a/A_i$, where $A_a$ is the area of telescope's aperture, stays the same as for the image in the restframe because the increased photon flux of the object due to apparent distance shortening is distributed over the lager area of the image. Our observable is distribution of radiation energy over the area of image or fluence F on photographic plate or CCD matrix accumulated during exposure time $\Delta t_e$; the illuminance is determined from the equation $I = F/\Delta t_{ex}$. In general, the illuminance I of image for any given distance to the object in the frame S is proportional to the brightness b that is diminished due to relativistic aberration together with redshift and time dilation by a factor $1/(1+z)^4$.

From the vantage point of a detector, the distance to the receding object is equal to D in the frame S at the moment of emission $t_e$ of the now-detected portion of light. As it was mentioned earlier, the spherical wavefront of radiation is spreading in all directions from the point in space where the object was at the moment $t_e$ and always stays centered at this point, therfore the flux on the detector should be proportional to $D^{-2}$. Relativistic aberration results in brightness/luminosity reduction by a factor $(1+z)^{-2}$ due to deflection of light rays; reddening of photons together with time dilation (elongation of time interval between the photons arriving to the detector) give another factor $(1+z)^{-2}$, totaling in luminosity reduction proportional to $1/D^2(1+z)^4$ and resulting in apparent luminosity distance increase $D_L = D(1+z)^2$. It means that with respect to the distance $D'$ in the restframe and, correspondingly, to the distance to the disk at the moment of detection the flux reduction factor is proportional to $1/D'^2(1+z)^2$ and the luminosity distance $D_L = D'(1+z)$. Thus, the effect of relativistic aberration "shifts" an object from its *distance from the apparent size* D to the distance $D(1+z)$ that corresponds to the distance to it at the moment of detection equal to the distance in the restframe $D'$, and the redshift together with time dilation shift it even farther by another factor $(1+z)$.

As for the non-resolvable objects, their geometric images are much smaller then the focal spots produced by the diffraction and optical aberrations of optical systems. From the point of view of optical system, these objects have no size and their visible angular size has no sense together with the effect related to the increase of their visible size. With no angular size there is no apparent angular size enlargement. For a pointlike object located on a line of sight, all original rays are directed out of the line of sight and, therefore, all the rays are further deflected from it by the relativistic aberration effect resulting in actual diminishing of apparent luminosity measured by the illuminance of focal spot with its size determined by diffraction and optical aberrations of optical system. The effect of relativistic aberration results in reduction of apparent luminosity of the receding source, in decreased flux of photons incident on the telescope's aperture, in decreased illuminance of the diffraction spot of an optical system, and in our judgment from this illuminance reduction about the apparent increase of the luminosity distance to the source by a factor $(1+z)$. Together with redshift and time dilation the luminosity distance $D_L = D(1+z)^2$.

3. RELATIVISTIC ABERRATION IN COSMIC SPACE

The effects of special relativity are obviously valid for the cosmic objects located in the detector's local space, i.e. for the objects with negligibly small cosmological redshift $z \ll 1$ associated with the expansion of the Universe. If objects move with relativistic velocity in the local space (such as jets emitted from active nuclei of nearby galaxies), their apparent brightness and luminosities are reduced or increased by a factor $\delta^4$ that corresponds to $(1+z)^4$, where z is their spectral redshift in the local space but not the cosmological redshift due to the Universe's expansion. In the case of jets, we know the distance to the host source from independent measurements, and this distance establishes the reference coordinate in the restframe for brightness/luminosity measurements of fast moving objects like jets from AGN.

The standard Big Bang cosmology considers the cosmological objects as moving away in stationary space (Newtonian concept) from any particular point in space, and their dynamics can be treated as occurring in Minkowski spacetime and obeying Friedmann equations. All the effects of special relativity are valid for the cosmological objects moving in stationary space therefore all the formulae of the previous section are also true for the cosmological redshifted objects.

From the Robertson-Walker metric written in the form $s^2 = c^2t^2 - R^2(t)[dr^2 + S_k^2(r)\,d\psi^2]$, where $S_k(r)$ is the function of the comoving coordinate r defined as sin r for k = 1, sinh r for k = -1, and r for k = 0 (Peacock 2002, 21), the transverse distance between two points δx in space located at the same cosmological redshift z is $R(t)S_k(r)d\psi$, where dψ is the angular difference between the points on the sky. The proper transverse size *dl* of a resolvable object is its comoving transverse size $S_k(r)d\psi$ times the scale factor at the time of emission: *dl* = $R_0 S_k(r)\,\delta\psi/(1+z)$, where $R_0$ is the scale factor at present. Therefore, like in special relativity, the apparent angular-diameter distance D = $R_0 S_k(r)/(1+z)$ is the distance at the moment of emission of now-detected photons. In the standard model, space does not stretch, the special relativity works as usual, and the wavefront after emission is centered on the point of emission at the distance D in space. It means that the apparent angular size is (1+z) times larger due to the relativistic aberration effect with respect to the restframe with z = 0 and the distance from the apparent size is (1+z) times shorter, i.e. by the same factor as for the angular-diameter distance D in the expanding Universe with respect to the distance related to the local space z = 0 determined by the RW metric. For an object of uniform brightness b(z) over its area A = $D^2$ δΩ, where δΩ is a solid angle subtended by an object and originated from a detector, the flux received by the detector from the solid angle δΩ on the sky subtended by the object *F* ≈ $b\int_0^A \frac{dA}{D^2} = b\int_0^{\delta\Omega} d\Omega = b\delta\Omega$, which is proportional to $(1+z)^{-2}$ (b is reduced by a factor $(1+z)^{-4}$ due to the combined effect of relativistic aberration, spectral redshift and time dilation and the angle $\delta\Omega$ is enlarged by a factor $(1+z)^2$ in comparison with the angular size in the restframe $z = 0$). The flux per unit solid angle on the sky $F/\delta\Omega = b(z)$ is reduced by a factor $(1+z)^{-4}$. If we use a radio telescope with its fixed angle $\Omega_a$ of detection, the measured radiation flux $S = (F/\delta\Omega)\delta\Omega_a = b(z) \delta\Omega_a$ so far as $\delta\Omega_a \leq \delta\Omega$; when $\delta\Omega_a > \delta\Omega$ the object becomes pointlike (unresolved) and

$S = L(z)A_a/D^2 = A_a(1+z)^{-4} D^{-2} \int_0^A b(0)dA$ independently of the actual angular size of the object.

In any case, the brightness/luminosity distance to a cosmological object evaluated from the measured flux $S$ is $D_L = (1+z)^2 D = (1+z) R_0 S_k(r)$, where $D = R(t_{em})S_k(r)$ is the distance at the moment of emission $t_{em}$ and $R_0 S_k(r)$ is the distance at the moment of detection which corresponds to its coordinate in the local space $z = 0$. In the case of optical telescopes, our observable is illuminance of image $E = (F a)/(f^2 \delta\Omega)$, where a is the area of telescope aperture and $f^2 \delta\Omega$ is the area of image, therefore the apparent brightness or luminosity distance is also $D_L = (1+z)^2 D = (1+z) R_0 S_k(r)$. As for the pointlike objects with their zero visible angular size the notion of flux from a solid angle on the sky becomes meaningless (formally, the flux from unit solid angle on the sky is infinite for finite luminosity of a pointlike object) and we have to consider luminosity $L$ reduced by a factor of $(1+z)^{-4}$ with respect to a stationary source at a distance D equal to the distance to the redshifted object at the moment of emission. As in the case of radio telescope measuring flux from an unresolved object, $S = L/D^2(1+z)^4$ and $D_L = (1+z)^2 R(t_{em})S_k(r) = (1+z) R_0 S_k(r)$. The relativistic aberration of redshifted objects in the expanding Universe 'sends' a redshifted object away from its distance at the moment of emission $R(t_{em})S_k(r)$ to its coordinate at the moment of detection $R_0 S_k(r)$, i.e. to its distance at the local space with $z = 0$; additionally, the redshift and the time dilation together 'send' it even farther by another factor $(1+z)$.

Two approaches to cosmology, the standard Big Bang model and the expanding space model, are currently coexisting accusing each other in fallacy and misconception (e.g., Peacock 2002, 30; Davis & Lineweaver 2004). It can be shown that the correction factor $(1+z)^{-4}$ for the measured flux from redshifted sources can be also obtained for the expanding space model. Consider a volume $dV = R^2 dR d\Omega$ between two spherical areas separated by dR

= c dt within a solid angle dΩ, containing N photons all moving in R-direction away from the source in the local space of a luminous redshifted pointlike object as shown in Fig. 4 (flat space is taken for simplicity). In a stationary space the angle δΩ would not change and neither would dR when the photons moved to the detector located at the distance D, therefore the observed volume $D^2 d\Omega dR$ would increase as $D^2$ and the flux of photons per unit area would decrease as $D^{-2}$. I an expending space, the photons move with the same velocity c to a detector while the space stretches omnidirectionally with the linear stretching factor proportional to the time dilation factor (1+z) for the moment of emission at a redshift z. It means that:

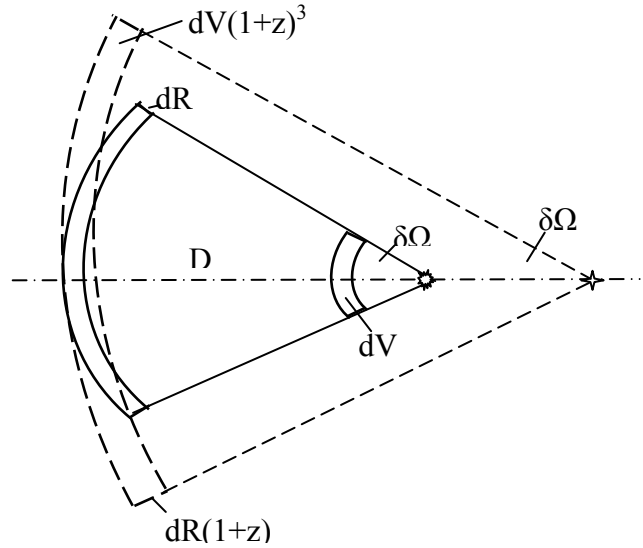

**Fig. 4.** Transformation of volume containing N photons between two spherical surfaces subtending a solid angle δΩ in the object's local space (solid lines) to the stretched space (dashed lines). An observer detects photons emitted by a source when it was at a distance D, however due to space stretching the volume is increased by a factor $(1+z)^3$ with corresponding reduction of photons' density while the wavefront is centered on the current distance of the source D(1+z). The account of wavelength stretching by another factor (1+z) yields the luminosity distance $D_L = D(1+z)^2 = R_0 r(1+z)$.

a) At the moment of detection, the considered spherical surfaces are separated by segment dR(1+z) due to space stretching that leads to (1+z)-fold reduction of photon linear density and to corresponding reduction of rate of their arrival to the detector, which is equal to the time dilation factor in special relativity.

b) Wavelength of each photon is enlarged by a factor (1+z) because each photon can be represented by a quasi-monochromatic electromagnetic wave with stretching distances between the wave's crests; the energy of photons is reduced correspondingly.

c) Finally, the areas of spherical surfaces of the volume containing the observed N photons are stretched by a factor $(1+z)^2$ in addition to their geometric enlargement proportional to $D^2$ in the nonstretching space. It leads to the additional reduction of photon flux per unit area at the detector's location by a factor $(1+z)^{-2}$.

d) Totally, the density of photons in the volume dV comoving with the photons is reduced by a factor $(1+z)^{-3}$, while the energy density is reduced by a factor $(1+z)^{-4}$ resulting in flux reduction by the same factor.

If one connects the edges of the stretched volume in Fig. 4 with the source at the distance D, the newly obtained solid angle will differ from the solid angle in the non-stretching restframe by a factor $(1+z)^2$, exactly the same as for the solid angle enlargement due to the effect of relativistic aberration in special relativity and in the standard Big Bang model. However, contrary to the special relativity theory and the standard Bing Bang model with non-stretching space, the form of the stretched wavefront (external surface of the stretched volume in Fig. 4) corresponds to the distance (radius) equal to the distance to the source at the moment of detection D(1+z). The spherical wavefront seen from a point of observation is always centered on the current position of the redshifted source of light at the moment of detection, and, therefore, the stretched spherical surface in Fig. 4 always subtends the same solid angle δΩ as in the object's local space. Photons in the expanding space always move along radii away from an object located at a current distance R(t)r, where r is the comoving coordinate and R(t) is the stretch scale of space, i.e. the inclination of rays with respect to the line-of-sight in the observer's local space are the same as in the object's local space. Contrary to the standard Big Bang model, there is no angular deflection of rays due to relativistic aberration; the reduction of photon flux at the detector's aperture is produced by the wavefront's transverse stretching in addition to redshift and time dilation on the condition that the detector's aperture stays constant while the space stretches, i.e. if the local measure (length unit) of physical bodies does not depend on the age of the Universe (sizes of atoms and intermolecular distances in solids are constant), so the detector's aperture would be the same if it existed at the cosmological time of emission of photons. If it does, i.e. if solid objects swell proportionally with the space stretching, the flux of photons captured by the detector's aperture is not influenced by the wavefront's transverse stretching, and only radial lengthening dR(1+z) of the volume dV in Fig.4 and redshift of photons count. If the first assumption is true, i.e. local spaces do not stretch with cosmological stretching, the reduction of flux captured by telescope aperture from a pointlike object is proportional to $(1+z)^{-4}$ like in the case of the standard Big Bang model and its luminosity distance $D_L = R_0 r(1+z)$; if the second hypothesis is true, the flux captured by the telescope aperture is reduced by a factor $(1+z)^{-2}$ and therefore the luminosity distance $D_L = R_0 r$, i.e. it is simply equal to the proper distance to the object at the moment of detection in the detector's local space $z = 0$ and no correction is required to scale the magnitudes in the z=0 template. The luminosity distance $D_L$ in the expanding space model is the same as in the standard Big Bang model, if the local space does not stretch with the age of the Universe and the local standard of length stays unchanged; otherwise, $D_L$ is $(1+z)$ times smaller. It is topologically difficult, if not impossible, to imagine the space that stretches at large but does not stretch locally in a finite vicinity of every point of the space. Either the space stretches everywhere together with the swelling local objects, including atoms and all microscopic bodies in general, which implies that the universal physical constants are the functions of cosmological time, or the expanding space model requires the assumption of local non-stretchability, which results in topological complexity and which can hardly withstand the test of Occam's razor. The observational evidences for luminosity distances to high-redshift galaxies (Sandage 1995) and supernovae (Schmidt 1998; Riess 1998; Perlmutter 1999) seem to comply with the standard Big Bang model.

4. CONCLUSION

Relativistic aberration of light emitted by cosmic objects is applicable equally to the objects moving with relativistic velocities in local space and to the cosmological objects in stationary space within the framework of the standard Big Bang model. In particular, the effect of relativistic aberration makes the sizable objects to appear closer to an observer, namely, at the *distance from the apparent size*, which corresponds to the angular-diameter distance $D_D$ obtained from the RW metric and which is $(1+z)$ times smaller then the distance $R_0 S_k(r)$ to the object in the restframe $z = 0$ at the moment of detection $t_0$, while the luminosity distance is $(1+z)$ times larger then $R_0 S_k(r)$. The same is also true for the pointlike objects. In the expending space model the luminosity distance $D_L = R_0 S_k(r)(1+z)$, if the space does not stretch locally and the local measures with physical constants do not depend on the age of the Universe; otherwise, the luminosity distance $D_L = R_0 S_k(r)$, i.e. (1+z) times smaller in comparison with the Big Bang model.